\documentclass[aps,prb,reprint,superscriptaddress,citeautoscript,amsmath,amssymb]{revtex4-2}

\usepackage{graphicx}
\usepackage{bm}
\usepackage{hyperref}
\hypersetup{hidelinks}

\begin{document}

\title{Annular Terahertz Near-Field of a Breathing Charged Vortex}

\author{Iogann Tolbatov}
\email[Author to whom correspondence should be addressed: ]{itolbatov@uniss.it}
\affiliation{Department of Chemical, Physical, Mathematical and Natural Sciences, University of Sassari, 07100 Sassari, Italy}

\author{Luca Salasnich}
\affiliation{Dipartimento di Fisica e Astronomia ``Galileo Galilei'' and Padua QTech Center, Universit\`a di Padova, Via Marzolo 8, 35131 Padova, Italy}
\affiliation{INFN Sezione di Padova, Via Marzolo 8, 35131 Padova, Italy}

\begin{abstract}
We study the radial electric field around a singly quantized vortex in a charged BCS superconductor, when the Higgs amplitude mode is driven by a uniform terahertz (THz) electromagnetic pulse. In a local-density approximation the pairing gap vanishes on the vortex axis at all times, so the particle density on the axis cannot oscillate, whereas the curvature of the density profile around the axis does. Starting from the electrostatic-lens formalism of our previous work, we show that this converts the electrically neutral Higgs oscillation into an alternating radial electric field that vanishes on the axis. When the electrostatic screening length is much shorter than the vortex core, we obtain the field in closed form: it has a ring-shaped profile, with a maximum at a distance of the order of the coherence length from the axis, and the Higgs drive modulates it by a fraction $2\eta$, where $\eta$ is the relative oscillation of the gap. For niobium nitride and $\eta=0.1$ the static field of the vortex is about $140$~V/m and its modulation about $30$~V/m, independently of the carrier density and of the dielectric constant. We also discuss how the thermodynamic factor entering the field changes toward the BCS-BEC crossover.
\end{abstract}

\maketitle

\section{Introduction}

Superconductors are characterized by a complex pairing order parameter \cite{Tinkham1996}. Its phase fluctuations give rise to the Goldstone mode, while its amplitude fluctuations give rise to the Higgs mode \cite{Pekker2015,Shimano2020}, which has been observed in THz experiments on conventional superconductors such as NbN \cite{Matsunaga2014,Matsunaga2017}. In a type-II superconductor a magnetic field penetrates in the form of quantized vortices \cite{Abrikosov1957,Blatter1994}: these are topological defects around which the phase of the order parameter winds by $2\pi$ and the pairing gap vanishes on the axis, recovering its bulk value over a coherence length \cite{Caroli1964}. A vortex core is also depleted of charge carriers and can therefore be electrically charged \cite{Khomskii1995}. A basic question is then how the Higgs mode, which is electrically neutral in a homogeneous system, can couple to electromagnetic fields. This question is central in THz spectroscopy, particularly in spatially inhomogeneous geometries where density gradients break translational invariance \cite{Sun2020}.

The penetration of electric fields in superconductors is usually modeled assuming a uniform bulk density. In a recent work \cite{TolbatovCondMat2026}, we showed that, once the density depletion near the core is taken into account, a static singly quantized charged vortex acts as a local electrostatic lens: the thermodynamic compressibility varies radially, and the scalar potential $\Phi$ of the electric field has the regular axial expansion
\begin{equation}
\Phi(r) = \Phi_0 + \Phi_2 r^2 + \Phi_4 r^4 + \dots ,
\label{eq:static_lens}
\end{equation}
where the quartic aberration coefficient $\Phi_4$ isolates the contribution of the equation of state. In the present work we extend this local-density framework to the dynamic regime.

In the static vortex the pairing field (the superconducting gap $\Delta$) vanishes at the center and heals to its bulk value $\Delta_0$ over the coherence length $\xi_{\rm GL}$. In a local-density description this static profile produces the density depletion and the electrostatic lens. A uniform THz pulse drives the Higgs mode, so that the gap oscillates and the density follows it locally. The density on the axis cannot oscillate, because the gap vanishes there at all times, but the curvature of the density profile around the axis does: the core ``breathes''. In this way the vortex converts the neutral Higgs oscillation into an alternating charge distribution, which generates an alternating radial electric field.

\section{Physical Setup and Dynamic Perturbation of the Electrostatic Lens}
\label{sec:dynamic}

We consider a thin film of a conventional $s$-wave type-II superconductor, such as NbN, in a weak static magnetic field $B$ perpendicular to the film. The field penetrates the film in the form of quantized vortices, each carrying one flux quantum $h/2e$; we consider a single, isolated vortex, pinned at the origin. The film is illuminated at normal incidence by a THz pulse, i.e.\ an electromagnetic wave of frequency $\omega_{\rm pump}$, polarized in the plane of the film. Its wavelength is many orders of magnitude larger than the size of the vortex core, so the pump field is spatially uniform around the vortex.

\begin{figure*}[t]
\centering
\includegraphics[width=0.95\textwidth]{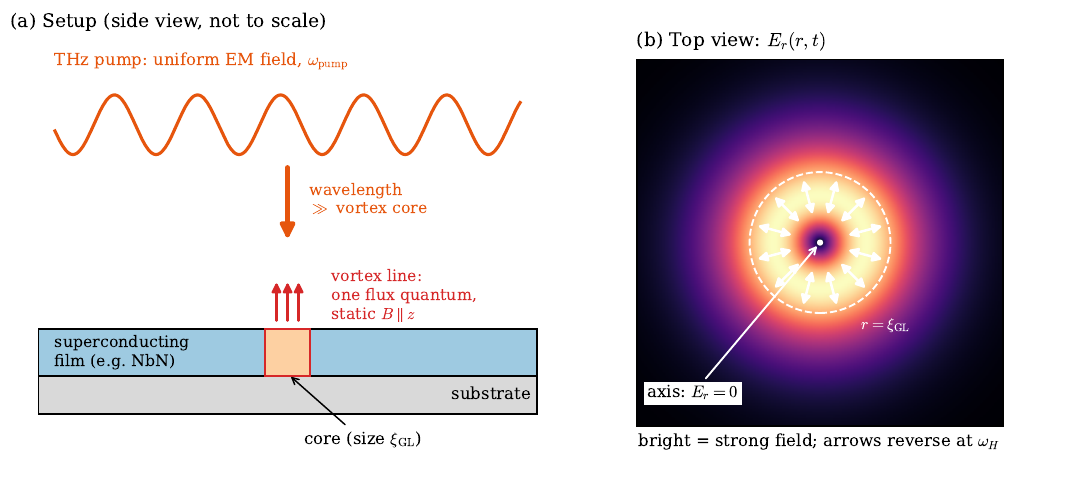}
\caption{Schematic representation of the physical setup. (a) A superconducting film in a static magnetic field $B$ perpendicular to it hosts a vortex (flux quantum $h/2e$). A THz pulse, whose wavelength is much larger than the core, illuminates the film uniformly. (b) Top view around the vortex: the radial electric field $E_r(r,t)$ (arrows) vanishes on the axis and is largest in a ring whose radius is of the order of $\xi_{\rm GL}$; it oscillates at the Higgs frequency $\omega_H$, with $\omega_H=2\omega_{\rm pump}$.}
\label{fig:setup}
\end{figure*}

Because the pump couples to the superconductor quadratically in the electromagnetic vector potential, it drives an oscillation of the gap amplitude at the frequency $2\omega_{\rm pump}=\omega_H\simeq2\Delta_0/\hbar$, where $\hbar$ is the reduced Planck constant and $\Delta_0$ is the bulk gap. This is the Higgs mode \cite{Shimano2020,Matsunaga2017}. In a homogeneous superconductor the oscillation changes the density by the same amount everywhere and produces no electric field, since the Higgs mode is electrically neutral. Near a vortex the gap is not uniform, so the density oscillates by a different amount at each distance from the axis; this non-uniform oscillating charge is the source of the electric field computed below. The field is radial, and oscillates in time at the frequency $\omega_H$ (Fig.~\ref{fig:setup}).

At $\omega_H$ the wavelength is also much larger than the region considered, so the field is quasi-static. Since the charge moves only radially and the geometry is invariant along the axis, there is no azimuthal magnetic field associated with it. We consider the electrostatic sector of the problem; the magnetic structure of the vortex, which varies on the scale of the London penetration depth, is not included.

The oscillation of the gap at the bulk Higgs frequency is the input of the model,
\begin{equation}
\Delta(r,t) = \bar\Delta(r) + \delta\Delta(r)\cos(\omega_H t),
\label{eq:gap}
\end{equation}
where $\bar\Delta(r)$ is the static profile and $\delta\Delta(r)$ the oscillation. In the deep-BCS limit the free Higgs mode lies at the edge of the two-quasiparticle continuum and decays as a power law in time, so a steady oscillation corresponds to continuous driving; the bound states of the core and the effects of finite-size confinement \cite{TolbatovMPLB2026} are not considered here.

For a singly quantized vortex the single-valuedness of the order parameter forces the amplitude to vanish on the axis at all times, $\Delta(0,t)=0$. Hence both the static gap and its oscillation grow linearly near the axis,
\begin{equation}
\bar\Delta(r) = \Delta_1 r + O(r^3),
\label{eq:gap_static}
\end{equation}
\begin{equation}
\delta\Delta(r) = \delta\Delta_1 r + O(r^3).
\label{eq:gap_dyn}
\end{equation}
Following the local-density extension of the phase-only Popov action, a formalism grounded in the coherent-state path integral approach \cite{Salasnich2026} and established in Ref.~\cite{TolbatovCondMat2026}, the quasi-static Poisson equation for the potential is
\begin{equation}
\frac{1}{r}\frac{\partial}{\partial r}\!\left(r\frac{\partial \Phi(r,t)}{\partial r}\right) - \kappa^2(r,t)\,\Phi(r,t) = -\frac{\rho(r,t)}{\epsilon},
\label{eq:poisson}
\end{equation}
where $\kappa^2(r,t)$ is the local electrostatic screening parameter fixed by the equation of state, and $\rho(r,t)=q\,[\,n(r,t)-\bar n_{\rm bg}]$ is the local charge density, $\bar n_{\rm bg}$ being the bulk density.

In the local-density approximation (LDA) the density follows the gap adiabatically, $n(r,t)=n(\mu,|\Delta(r,t)|^2)$, which is the leading-order quasi-static description of the density response at $\omega_H$. Equating the linearized response to a near-axis expansion,
\begin{equation}
n(r,t)=n_c(t)+n_2(t)\,r^2+O(r^4),
\label{eq:density_expansion}
\end{equation}
gives a thermodynamic constraint: since the gap vanishes on the axis at all times, the central density cannot oscillate,
\begin{equation}
\delta n_c = 0 .
\label{eq:dnc}
\end{equation}
The Higgs mode modulates only the curvature of the density profile,
\begin{equation}
\delta n_2 = 2\chi_c\,\Delta_1\,\delta\Delta_1 ,
\label{eq:dn2}
\end{equation}
where $\chi_c=(\partial n/\partial|\Delta|^2)_\mu$ is the susceptibility of the density to the gap at the core center.

Integrating Eq.~\eqref{eq:poisson} over the plane, for a field that decays at large $r$, gives $\int(\rho-\epsilon\kappa^2\Phi)\,d^2r=0$: the total charge per unit length, i.e.\ the bare charge plus the induced screening charge $-\epsilon\kappa^2\Phi$, vanishes identically. This compensation takes place at $r\gtrsim\xi_{\rm GL}$ and is encoded in the boundary value $\delta\Phi_0$ of the potential.

We expand the potential, the screening parameter and the charge density in a static (dc) part and a linear dynamic (ac) part oscillating at $\omega_H$:
\begin{align}
\Phi(r,t) &= [\bar\Phi_0+\delta\Phi_0(t)] + [\bar\Phi_2+\delta\Phi_2(t)]\,r^2 \nonumber\\
&\quad + [\bar\Phi_4+\delta\Phi_4(t)]\,r^4, \label{eq:Phi_exp}
\end{align}
\begin{equation}
\kappa^2(r,t) = [\bar\kappa_0^2+\delta\kappa_0^2(t)] + [\bar C+\delta C(t)]\,r^2,
\label{eq:kappa_exp}
\end{equation}
\begin{equation}
\rho(r,t) = [\bar\rho_0+\delta\rho_0(t)] + [\bar\rho_2+\delta\rho_2(t)]\,r^2.
\label{eq:rho_exp}
\end{equation}
The time-independent part reproduces the steady-state electrostatic lens of Ref.~\cite{TolbatovCondMat2026}; in particular the static coefficients are
\begin{equation}
\bar\Phi_2 = \frac{\bar\kappa_0^2}{4}\bar\Phi_0-\frac{\bar\rho_0}{4\epsilon},
\label{eq:Phi2_static}
\end{equation}
\begin{equation}
\bar\Phi_4 = \frac{\bar\kappa_0^4}{64}\bar\Phi_0-\frac{\bar\kappa_0^2}{64\epsilon}\bar\rho_0+\frac{\bar C}{16}\bar\Phi_0-\frac{\bar\rho_2}{16\epsilon}.
\label{eq:Phi4_static}
\end{equation}
Applying a linear variation $\delta$ to these static solutions, and using $\delta n_c=0$, we obtain
\begin{equation}
\delta\rho_0=0,
\label{eq:drho0}
\end{equation}
\begin{equation}
\delta\kappa_0^2=0 ,
\label{eq:dkappa0}
\end{equation}
so that the whole coupling to the Higgs mode enters at order $r^2$, through the bare charge fluctuation $\delta\rho_2=q\,\delta n_2$ and the oscillation of the compressibility gradient $\delta C=\kappa_c^{2\prime}\,\delta n_2$. At order $r^0$ the dynamic paraxial lens depends only on the boundary value,
\begin{equation}
\delta\Phi_2=\frac{\bar\kappa_0^2}{4}\,\delta\Phi_0 ,
\label{eq:dPhi2}
\end{equation}
and at order $r^2$ we obtain
\begin{equation}
\delta\Phi_4=\frac{\bar\kappa_0^4}{64}\delta\Phi_0+\frac{\bar C}{16}\delta\Phi_0+\frac{\bar\Phi_0}{16}\,\delta C-\frac{\delta\rho_2}{16\epsilon}.
\label{eq:dPhi4}
\end{equation}
The terms proportional to $\delta\Phi_0$ are the passive ac screening response; the last two terms are the active contribution, driven by the oscillating density curvature $\delta n_2$. The value of $\delta\Phi_0$ is fixed by matching the core to the bulk, which mixes the two contributions.

\section{Electrostatic Screening in the Deep-BCS Regime}
\label{sec:deepBCS}

We now specialize to a charged Fermi superfluid ($q=-e$) in the weakly interacting deep-BCS limit. The chemical potential of the three-dimensional Fermi gas is $\mu=c_\mu n^{2/3}$, with $c_\mu=(3\pi^2)^{2/3}\hbar^2/(2m)$, $m$ the fermion mass and $n$ the number density. The local compressibility is then $\partial n/\partial\mu\propto n^{1/3}$, so that the screening parameter is
\begin{equation}
\kappa^2(n)=\alpha_\kappa\, n^{1/3},
\label{eq:kappa_n}
\end{equation}
with
\begin{equation}
\alpha_\kappa=\frac{3q^2}{2\epsilon\,c_\mu}.
\label{eq:alpha_kappa}
\end{equation}
The central screening parameter and the compressibility gradient are
\begin{equation}
\bar\kappa_0^2=\alpha_\kappa\bar n_c^{1/3},
\label{eq:kappa_c}
\end{equation}
\begin{equation}
\kappa_c^{2\prime}=\tfrac13\,\alpha_\kappa\,\bar n_c^{-2/3},
\label{eq:kappa_cp}
\end{equation}
and therefore
\begin{equation}
\bar C=\tfrac13\,\alpha_\kappa\,\bar n_c^{-2/3}\,\bar n_2,
\label{eq:C}
\end{equation}
\begin{equation}
\delta C=\tfrac13\,\alpha_\kappa\,\bar n_c^{-2/3}\,\delta n_2 .
\label{eq:dC}
\end{equation}
Inserting these into Eq.~\eqref{eq:dPhi4} gives
\begin{align}
\delta\Phi_4 &= \left[\frac{\alpha_\kappa^2\bar n_c^{2/3}}{64}+\frac{\alpha_\kappa\bar n_c^{-2/3}\bar n_2}{48}\right]\delta\Phi_0 \nonumber\\
&\quad+\frac1{16}\left[\frac13\alpha_\kappa\bar\Phi_0\,\bar n_c^{-2/3}-\frac q\epsilon\right]\delta n_2 ,
\label{eq:dPhi4_deep}
\end{align}
where the first bracket is the passive screening of the boundary condition and the second is the active term.

In the Thomas-Fermi regime ($\bar\kappa_0\xi_{\rm GL}\gg1$) the static core potential reduces to
\begin{equation}
\bar\Phi_0=\frac{q}{\epsilon\bar\kappa_0^2}\,(\bar n_c-\bar n_{\rm bg}),
\label{eq:Phi0_TF}
\end{equation}
and the active term becomes
\begin{equation}
\delta\Phi_{4,\rm active}=-\frac{q}{16\epsilon}\left[\frac23+\frac13\frac{\bar n_{\rm bg}}{\bar n_c}\right]\delta n_2 .
\label{eq:active_factor}
\end{equation}
In the weakly interacting limit the core depletion is small ($\bar n_c\to\bar n_{\rm bg}$) and the factor in brackets tends to unity; it grows if the core is more strongly depleted (Sec.~\ref{sec:crossover}).

\section{The Radial Electric Field}
\label{sec:field}

In the quasi-static limit $E_r=-\partial_r\Phi$, and the dynamic part of the field is
\begin{equation}
\delta E_r(r)=-2\,\delta\Phi_2\,r-4\,\delta\Phi_4\,r^3 .
\label{eq:Er_gen}
\end{equation}
Using Eqs.~\eqref{eq:dPhi2} and \eqref{eq:dPhi4_deep},
\begin{align}
\delta E_r(r)&=-\frac{\bar\kappa_0^2}{2}\,\delta\Phi_0\,r-4F(\bar n_c,\bar n_2)\,\delta\Phi_0\,r^3 \nonumber\\
&\quad-\frac14\left[\frac13\alpha_\kappa\bar\Phi_0\bar n_c^{-2/3}-\frac q\epsilon\right]\delta n_2\,r^3,
\label{eq:Er_deep}
\end{align}
with $F=\alpha_\kappa^2\bar n_c^{2/3}/64+\alpha_\kappa\bar n_c^{-2/3}\bar n_2/48$. The field vanishes on the axis and has no divergence there, so the core does not behave as a pulsing point charge. The coefficients contain $\bar\kappa_0^2$, $\bar\kappa_0^4$, \dots, so this paraxial expansion is an expansion in $(\kappa_0 r)^2$ and describes distances $r\ll\kappa_0^{-1}$.

\begin{figure*}[!t]
\centering
\includegraphics[width=0.92\textwidth]{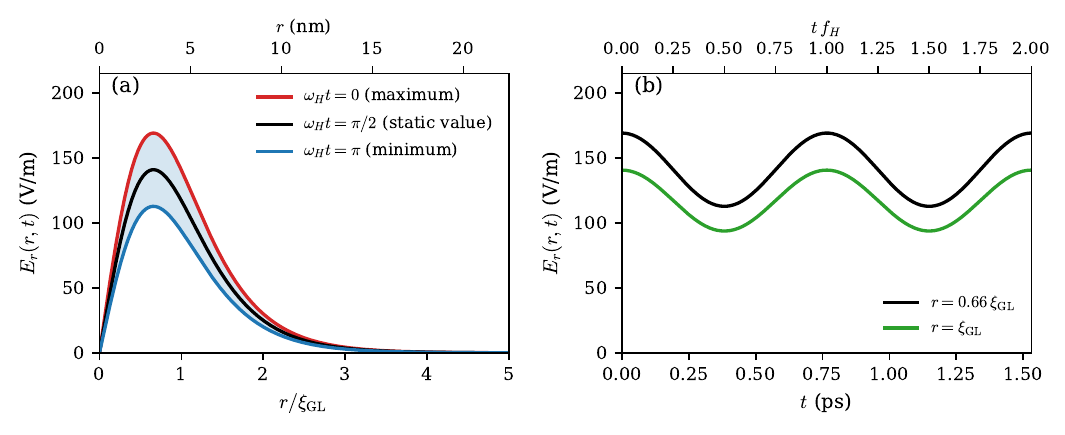}
\caption{Radial electric field of the vortex for the NbN parameters of Table~\ref{tab:params} and $\eta=0.1$. (a) $E_r$ as a function of the distance $r$ from the axis at three instants of the Higgs oscillation: at $\omega_Ht=0$ (maximum), $\pi/2$ (static value) and $\pi$ (minimum); the shaded band is the range swept during one period. The field vanishes on the axis and has a ring-shaped maximum at $r\simeq0.66\,\xi_{\rm GL}$. (b) $E_r$ as a function of time at two distances from the axis, over two periods of the oscillation ($f_H=1.31$~THz).}
\label{fig:fields}
\end{figure*}

For distances $r\gg\kappa_0^{-1}$ the potential follows the local density. Neglecting the gradient term in Eq.~\eqref{eq:poisson} (the corrections are $O[(\kappa_0\xi_{\rm GL})^{-2}]$) gives $\Phi\simeq q[n-\bar n_{\rm bg}]/[\epsilon\kappa^2(n)]$, and linearizing around the static profile, with $\kappa^2\propto n^\gamma$ and $\gamma=1/3$,
\begin{equation}
\delta\Phi(r)=\frac{q\,\delta n(r)}{\epsilon\,\bar\kappa^2(r)}\left[(1-\gamma)+\gamma\frac{\bar n_{\rm bg}}{\bar n(r)}\right],
\label{eq:TF_local}
\end{equation}
which contains the same thermodynamic factor as Eq.~\eqref{eq:active_factor}. In the deep-BCS regime this factor is unity.

To evaluate the field we take the simplest vortex profile consistent with Eqs.~\eqref{eq:gap_static} and \eqref{eq:gap_dyn},
\begin{equation}
\bar\Delta(r)=\Delta_0\tanh\frac{r}{\xi_{\rm GL}},
\label{eq:profile_gap}
\end{equation}
so that $\Delta_1=\Delta_0/\xi_{\rm GL}$, and a uniform relative modulation $\delta\Delta(r)=\eta\,\bar\Delta(r)$, with $\eta=\delta\Delta_0/\Delta_0$ the relative amplitude of the bulk gap oscillation. Taking $\chi_c$ as constant, the density is
\begin{equation}
n(r,t)=\bar n_{\rm bg}-\chi_c\Delta_0^2\left[1-(1+\eta\cos\omega_Ht)^2\tanh^2\frac{r}{\xi_{\rm GL}}\right],
\label{eq:n_profile}
\end{equation}
which gives $\delta n_c=0$ and $\delta n_2=2\eta\chi_c\Delta_0^2/\xi_{\rm GL}^2$, in agreement with Eq.~\eqref{eq:dn2}. The density far from the core oscillates uniformly; this produces a spatially uniform potential and no field. Using the dimensional estimate $\chi_c\simeq\bar n_{\rm bg}/E_F^2$ (up to a factor of order unity), with $E_F$ the Fermi energy, and $\kappa_0^2=\tfrac32\,e^2\bar n_{\rm bg}/(\epsilon E_F)$ (from $\partial n/\partial\mu=3n/2E_F$), the density and the dielectric constant cancel in Eq.~\eqref{eq:TF_local}. To first order in $\eta$ the radial field $E_r=-\partial_r\Phi$ is
\begin{equation}
E_r(r,t)=E_0\,{\rm sech}^2\!\frac{r}{\xi_{\rm GL}}\,\tanh\frac{r}{\xi_{\rm GL}}\,\bigl[1+2\eta\cos(\omega_Ht)\bigr],
\label{eq:E_TF}
\end{equation}
with the field scale
\begin{equation}
E_0=\frac43\,\frac{\Delta_0^2}{e\,E_F\,\xi_{\rm GL}} .
\label{eq:E0}
\end{equation}
The static part of Eq.~\eqref{eq:E_TF} is the charge-depletion field of the vortex \cite{TolbatovCondMat2026}, and the Higgs drive modulates it by the fraction $2\eta$. The field is ring-shaped: it is linear in $r$ near the axis (as for any smooth axisymmetric potential), has its maximum at $\tanh(r/\xi_{\rm GL})=1/\sqrt3$, i.e.\ $r_{\max}\simeq0.66\,\xi_{\rm GL}$, and decays exponentially at large $r$. The static value at the maximum is
\begin{equation}
E_{r,\max}=\frac{8}{9\sqrt3}\,\frac{\Delta_0^2}{e\,E_F\,\xi_{\rm GL}} ,
\label{eq:Emax}
\end{equation}
and the amplitude of its oscillation is $2\eta E_{r,\max}$.

We use the parameters of epitaxial NbN, a standard deep-BCS superconductor (Table~\ref{tab:params}). The clean-limit length is $\xi_0=\hbar v_F/\Delta_0$ \cite{Sun2020}, but epitaxial NbN films are in the dirty limit ($l\ll\xi_0$), so the relevant scale is the Ginzburg-Landau length $\xi_{\rm GL}=4.5$~nm \cite{Chockalingam2008}. The gap $\Delta_0\simeq2.7$~meV is an empirical input \cite{Chockalingam2008}; it fixes $\omega_H\simeq2\Delta_0/\hbar\simeq8.2\times10^{12}$~s$^{-1}$, i.e.\ $f_H\simeq1.31$~THz \cite{Shimano2020,Matsunaga2017}. The carrier density $\bar n_{\rm bg}=6.5\times10^{28}$~m$^{-3}$ is from Hall measurements \cite{Chockalingam2008}; for a free-electron gas $E_F=\hbar^2(3\pi^2\bar n_{\rm bg})^{2/3}/2m_e\simeq5.9$~eV and $\Delta_0/E_F\simeq4.6\times10^{-4}$. The factor $(\Delta_0/E_F)^2$, which expresses the weak coupling between the gap amplitude and the density (particle-hole symmetry), enters through the estimate of $\chi_c$ \cite{Sun2020}. Gap modulations of tens of percent have been reported under intense THz driving \cite{Cui2019}, while weaker driving leaves the gap weakly perturbed \cite{Huang2025}; we take $\eta=0.1$. With $\epsilon=\epsilon_0$ the screening length is $\kappa_0^{-1}\simeq0.58$~\AA, so $\kappa_0\xi_{\rm GL}\simeq78$ and the Thomas-Fermi regime applies for $r\gtrsim0.1\,\xi_{\rm GL}$, where the local description of screening holds ($r\gg k_F^{-1}\simeq0.8$~\AA).

\begin{table}[t]
\caption{Parameters for epitaxial NbN used in the estimates ($\epsilon=\epsilon_0$, free-electron mass).}
\label{tab:params}
\small
\begin{tabular}{lcc}
\hline\hline
Parameter & Symbol & Value\\
\hline
Background density & $\bar n_{\rm bg}$ & $6.5\times10^{28}$ m$^{-3}$\\
Fermi energy & $E_F$ & $5.9$ eV\\
Superconducting gap & $\Delta_0$ & $2.7$ meV\\
Gap-to-Fermi ratio & $\Delta_0/E_F$ & $4.6\times10^{-4}$\\
Coherence length & $\xi_{\rm GL}$ & $4.5$ nm\\
Higgs frequency & $f_H=\omega_H/2\pi$ & $1.31$ THz\\
Rel.\ gap modulation & $\eta$ & $0.1$\\
Screening length & $\kappa_0^{-1}$ & $0.58$ \AA\\
Inv.\ Fermi wavevector & $k_F^{-1}$ & $0.80$ \AA\\
Thomas-Fermi par. & $\kappa_0\xi_{\rm GL}$ & $78$\\
\hline\hline
\end{tabular}
\end{table}

We solved the radial Poisson equation, Eq.~\eqref{eq:poisson}, numerically for the profile of Eq.~\eqref{eq:n_profile}, with regularity on the axis and decay at large $r$. The numerical solution agrees with Eq.~\eqref{eq:E_TF} to better than $0.2\%$ for $0.005\lesssim r/\xi_{\rm GL}\lesssim6$. The result is shown in Fig.~\ref{fig:fields}. The static field reaches $141$~V/m at $r\simeq0.66\,\xi_{\rm GL}\simeq3$~nm, and the Higgs drive modulates it by $\pm20\%$, i.e.\ by an oscillating component of amplitude $28$~V/m ($0.28$~V/cm) at $f_H=1.31$~THz. At $r=\xi_{\rm GL}$ the static field is $117$~V/m and the oscillating amplitude $23$~V/m. The corresponding potential difference across the core is of order $E_{r,\max}\xi_{\rm GL}\sim0.6\,\mu$V.

The oscillating component is proportional to $\eta$ in the linear regime, and both the static and the oscillating parts are proportional to $\chi_c$. For a pure modulation of the core width, $\xi_{\rm GL}\to\xi_{\rm GL}(1+\eta)$, in which the bulk gap does not oscillate, the same calculation gives an oscillating component of maximum $19$~V/m at $r\simeq0.42\,\xi_{\rm GL}$. In the Thomas-Fermi regime the dielectric constant and the carrier density cancel in Eqs.~\eqref{eq:E_TF}--\eqref{eq:Emax}; they enter only through the condition $\kappa_0\xi_{\rm GL}\gg1$, with $\kappa_0\propto\epsilon^{-1/2}$.

\section{BCS-BEC Crossover}
\label{sec:crossover}

In the deep-BCS limit the core depletion is small and the thermodynamic factor in Eqs.~\eqref{eq:active_factor} and \eqref{eq:TF_local} is unity. The factor scales with $\bar n_{\rm bg}/\bar n_c$, so it increases when the core is more strongly depleted, as in the BCS-BEC crossover. For strongly depleted cores the local compressibility must be obtained from the full equation of state $n(\mu,\Delta)$ across the 3D crossover \cite{Manini2004}, instead of $\kappa^2\propto n^{1/3}$. As the coupling parameter $\Delta_0/E_F$ increases, the Higgs resonance is pushed into the pair-breaking continuum, where the complex poles of the fluctuation propagator show that it survives with a finite damping rate \cite{Kurkjian2019,Kurkjian2020}, also in charged superfluids across the crossover \cite{Klimin2023}; coherent Higgs oscillations have been observed even at unitarity \cite{Dyke2024}. The increase of the thermodynamic factor and the damping of the Higgs mode act in opposite directions, and their balance can be addressed with a self-consistent solution of the Bogoliubov-de~Gennes equations beyond the LDA.

\section{Conclusions}

We have studied, in a local-density approximation, the radial electric field of a charged vortex whose Higgs amplitude mode is driven by a uniform THz pulse. Because the superconducting gap vanishes on the vortex axis, the central carrier density cannot oscillate, but the curvature of the density profile does. This spatial variation converts the electrically neutral Higgs oscillation into an alternating radial electric field. In the Thomas-Fermi regime, the field takes a simple closed form (Eq.~\eqref{eq:E_TF}): it features a ring-shaped profile whose static value is modulated by the fraction $2\eta$ at the Higgs frequency, independent of the background carrier density and the dielectric constant. 

For parameters typical of epitaxial NbN, the resulting oscillating component is on the order of tens of V/m. While macroscopic far-field radiation is absent due to the vanishing total induced charge, these localized near-field oscillations could potentially be probed by scattering-type scanning near-field optical microscopy (s-SNOM) \cite{Chen2019, Kim2023} or scanning tunneling microscopy coupled with THz excitation (THz-STM) \cite{Cocker2013, Xu2026}, which are highly sensitive to local charge dynamics. Furthermore, because the thermodynamic factor driving the active screening term increases for more strongly depleted cores, the dynamic electric field is expected to be significantly enhanced as the system is tuned toward the BCS-BEC crossover. This mechanism thus provides a purely electromagnetic pathway to dynamically probe the equation of state and the localized Higgs mode inside topological defects.

\begin{acknowledgments}
L.S. is partially supported by the Project ``Frontiere Quantistiche'' (Dipartimenti di Eccellenza) of the Italian Ministry of University and Research and by ``Iniziativa Specifica Quantum'' of INFN.
\end{acknowledgments}


\begin{thebibliography}{99}

\bibitem{Tinkham1996}
M. Tinkham,
\textit{Introduction to Superconductivity}, 2nd ed.
(McGraw-Hill, New York, 1996).

\bibitem{Pekker2015}
D. Pekker and C. M. Varma,
\textit{Amplitude/Higgs modes in condensed matter physics},
Annu. Rev. Condens. Matter Phys. \textbf{6}, 269 (2015).

\bibitem{Shimano2020}
R. Shimano and N. Tsuji,
\textit{Higgs mode in superconductors},
Annu. Rev. Condens. Matter Phys. \textbf{11}, 103 (2020).

\bibitem{Matsunaga2014}
R. Matsunaga, Y. I. Hamada, K. Makise, Y. Uzawa, H. Terai, Z. Wang, and R. Shimano,
\textit{Light-induced collective pseudospin precession resonating with Higgs mode in a superconductor},
Science \textbf{345}, 1145 (2014).

\bibitem{Matsunaga2017}
R. Matsunaga, N. Tsuji, K. Makise, H. Terai, H. Aoki, and R. Shimano,
\textit{Polarization-resolved terahertz third-harmonic generation in a single-crystal superconductor NbN: Dominance of the Higgs mode beyond the BCS approximation},
Phys. Rev. B \textbf{96}, 020505(R) (2017).

\bibitem{Abrikosov1957}
A. A. Abrikosov,
\textit{On the magnetic properties of superconductors of the second group},
Sov. Phys. JETP \textbf{5}, 1174 (1957).

\bibitem{Blatter1994}
G. Blatter, M. V. Feigel'man, V. B. Geshkenbein, A. I. Larkin, and V. M. Vinokur,
\textit{Vortices in high-temperature superconductors},
Rev. Mod. Phys. \textbf{66}, 1125 (1994).

\bibitem{Caroli1964}
C. Caroli, P. G. de Gennes, and J. Matricon,
\textit{Bound fermion states on a vortex line in a type II superconductor},
Phys. Lett. \textbf{9}, 307 (1964).

\bibitem{Khomskii1995}
D. I. Khomskii and A. Freimuth,
\textit{Charged vortices in high temperature superconductors},
Phys. Rev. Lett. \textbf{75}, 1384 (1995).

\bibitem{Sun2020}
Z. Sun, M. M. Fogler, D. N. Basov, and A. J. Millis,
\textit{Collective modes and terahertz near-field response of superconductors},
Phys. Rev. Research \textbf{2}, 023413 (2020).

\bibitem{TolbatovCondMat2026}
I. Tolbatov and L. Salasnich,
\textit{Electrostatic lens of a charged vortex},
Condens. Matter \textbf{11}, 35 (2026).

\bibitem{TolbatovMPLB2026}
I. Tolbatov and L. Salasnich,
\textit{Curvature-induced shifts of Higgs and Goldstone modes in 2D superconducting shells},
Mod. Phys. Lett. B (Accepted, 2026).

\bibitem{Salasnich2026}
L. Salasnich and C. Vianello,
\textit{Coherent-state path integrals in quantum thermodynamics},
SciPost Phys. Lect. Notes \textbf{117} (2026).

\bibitem{Chockalingam2008}
S. P. Chockalingam, M. Chand, J. Jesudasan, V. Tripathi, and P. Raychaudhuri,
\textit{Superconducting properties and Hall effect of epitaxial NbN thin films},
Phys. Rev. B \textbf{77}, 214503 (2008).

\bibitem{Cui2019}
T. Cui, X. Yang, C. Vaswani, J. Wang, R. M. Fernandes, and P. P. Orth,
\textit{Impact of damping on the superconducting gap dynamics induced by intense terahertz pulses},
Phys. Rev. B \textbf{100}, 054504 (2019).

\bibitem{Huang2025}
C. Huang \textit{et al.},
\textit{Discovery of an unconventional quantum echo by interference of Higgs coherence},
Sci. Adv. \textbf{11}, eads8740 (2025).

\bibitem{Manini2004}
N. Manini and L. Salasnich,
\textit{Bulk and collective properties of a dilute Fermi gas in the BCS-BEC crossover},
Phys. Rev. A \textbf{71}, 033625 (2005).

\bibitem{Kurkjian2019}
H. Kurkjian, S. N. Klimin, J. Tempere, and Y. Castin,
\textit{Pair-breaking collective branch in BCS superconductors and superfluid Fermi gases},
Phys. Rev. Lett. \textbf{122}, 093403 (2019).

\bibitem{Kurkjian2020}
H. Kurkjian, J. Tempere, and S. N. Klimin,
\textit{Linear response of a superfluid Fermi gas inside its pair-breaking continuum},
Sci. Rep. \textbf{10}, 11591 (2020).

\bibitem{Klimin2023}
S. N. Klimin, J. Tempere, T. Repplinger, and H. Kurkjian,
\textit{Collective excitations of a charged Fermi superfluid in the BCS-BEC crossover},
New J. Phys. \textbf{25}, 063011 (2023).

\bibitem{Dyke2024}
P. Dyke, S. Musolino, H. Kurkjian, D. J. M. Ahmed-Braun, A. Pennings, I. Herrera, S. Hoinka, S. J. J. M. F. Kokkelmans, V. E. Colussi, and C. J. Vale,
\textit{Higgs oscillations in a unitary Fermi superfluid},
Phys. Rev. Lett. \textbf{132}, 223402 (2024).

\bibitem{Chen2019}
X. Chen, D. Hu, R. Mescall, G. You, D. N. Basov, Q. Dai, and M. Liu,
\textit{Modern scattering-type scanning near-field optical microscopy for advanced material research},
Adv. Mater. \textbf{31}, 1804774 (2019).

\bibitem{Kim2023}
R. H. J. Kim, J.-M. Park, S. J. Haeuser, L. Luo, and J. Wang,
\textit{A sub-2 Kelvin cryogenic magneto-terahertz scattering-type scanning near-field optical microscope (cm-THz-sSNOM)},
Rev. Sci. Instrum. \textbf{94}, 043702 (2023).

\bibitem{Cocker2013}
T. L. Cocker, V. Jelic, M. Gupta, S. J. Molesky, J. A. J. Burgess, G. De Los Reyes, L. V. Titova, Y. Y. Tsui, M. R. Freeman, and F. A. Hegmann,
\textit{An ultrafast terahertz scanning tunnelling microscope},
Nat. Photonics \textbf{7}, 620 (2013).

\bibitem{Xu2026}
J. Xu, Y. Xia, Q. Lin, S. Tan, B. Wang, and J.-X. Yin,
\textit{Ultrahigh spatiotemporal resolution terahertz scanning tunnelling microscopy},
Nat. Rev. Phys. \textbf{8}, 508 (2026).

\end{thebibliography}
\end{document}